\documentclass{article}
\usepackage[T1]{fontenc} 
\usepackage[utf8]{inputenc} 
\usepackage{ismir,amsmath,cite,url}
\usepackage{graphicx}
\usepackage{color}
\usepackage{amssymb}
\usepackage{tabularx} 
\usepackage{booktabs} 
\usepackage{lineno}
\usepackage{multirow}

\title{MelodyT5: A Unified Score-to-Score Transformer for Symbolic Music Processing}

\multauthor
{Shangda Wu$^{1,\sharp}$\thanks{$^\sharp$ These authors contributed equally.} \hspace{1cm} Yashan Wang$^{1,\sharp}$ \hspace{1cm} Xiaobing Li$^1$ \hspace{1cm} Feng Yu $^1$ \hspace{1cm} Maosong Sun$^{1,2,\flat}$\thanks{$^\flat$ Corresponding author.}}
{ $^1$ Central Conservatory of Music, China \hspace{1cm} $^2$ Tsinghua University, China\\
{\tt\small \{shangda, alexis\_wang\}@mail.ccom.edu.cn, \ sms@tsinghua.edu.cn} \\
{\small\url{https://github.com/sanderwood/melodyt5}}
}
\def\authorname{S. Wu, Y. Wang, X. Li, F. Yu, and M. Sun}

\begin{document}

\maketitle
\begin{abstract}
In the domain of symbolic music research, the progress of developing scalable systems has been notably hindered by the scarcity of available training data and the demand for models tailored to specific tasks. To address these issues, we propose MelodyT5, a novel unified framework that leverages an encoder-decoder architecture tailored for symbolic music processing in ABC notation. This framework challenges the conventional task-specific approach, considering various symbolic music tasks as score-to-score transformations. Consequently, it integrates seven melody-centric tasks, from generation to harmonization and segmentation, within a single model. Pre-trained on MelodyHub, a newly curated collection featuring over 261K unique melodies encoded in ABC notation and encompassing more than one million task instances, MelodyT5 demonstrates superior performance in symbolic music processing via multi-task transfer learning. Our findings highlight the efficacy of multi-task transfer learning in symbolic music processing, particularly for data-scarce tasks, challenging the prevailing task-specific paradigms and offering a comprehensive dataset and framework for future explorations in this domain.
\end{abstract}
\section{Introduction}\label{sec:introduction}
In the field of artificial intelligence, symbolic music processing—including the analysis and generation of musical scores—presents a unique challenge that merges musical creativity with computational complexity. Symbolic music, which represents musical information with discrete symbols rather than continuous audio signals, facilitates the precise manipulation and analysis of elements such as melody, harmony, and rhythm. Historically, the application of AI in this area has sought not only to mimic the creative process of human composers \cite{DBLP:conf/ismir/LiuDCZLYS22, DBLP:conf/ismir/Lu0YQZL22, DBLP:conf/ismir/MinJXZ23, wu2023exploring} but also to uncover the underlying patterns of musical composition \cite{DBLP:conf/acl/ZengTWJQL21, DBLP:conf/ismir/0008X21, DBLP:conf/ismir/WuY0S23}.

Despite significant progress, the field still faces persistent limitations. One notable challenge is the prevalence of task-specific models \cite{zhang2021symbolic, DBLP:journals/access/ChoiPHJP21, DBLP:conf/icassp/WuLS23, DBLP:journals/ejasmp/WuYWLS24}. These models offer benefits for specific applications but lack adaptability to the broader spectrum of symbolic music processing. This fragmentation is further compounded by the scarcity of annotated datasets \cite{DBLP:conf/ismir/00080JZXDX20, DBLP:conf/ismir/HsiaoHC023, DBLP:journals/tismir/ZhangZLYS23}, which serve as the lifeblood of deep learning models. Unlike other domains where data may be abundant and easy-to-collect, annotated symbolic music datasets are both rare and costly to produce. Without access to ample and diverse data, models struggle to generalize and may exhibit biases or limitations \cite{DBLP:journals/tismir/HolzapfelSC18} in their analysis and generation of symbolic music.

In addressing the challenges inherent to symbolic music processing, insights from the Natural Language Processing (NLP) domain offer a promising avenue for advancement. Techniques such as transfer learning \cite{DBLP:conf/acl/RuderH18, DBLP:conf/naacl/RuderPSW19, DBLP:journals/tacl/ArtetxeS19} and multi-task learning \cite{DBLP:conf/acl/LiuHCG19, DBLP:conf/icml/SongTQLL19, DBLP:conf/eacl/ZhangYYGJ23} have played a pivotal role in advancing NLP by promoting the transfer of knowledge from pre-trained language models and exploiting common patterns across various tasks. Prominent models like GPT \cite{radford2018improving}, BERT \cite{DBLP:conf/naacl/DevlinCLT19}, and T5 \cite{DBLP:journals/jmlr/RaffelSRLNMZLL20} demonstrate the efficacy of these strategies in understanding and generating language across diverse contexts. Notably, the T5 model, with its text-to-text framework, mirrors the conceptual shift necessary for symbolic music by treating all tasks as variations of converting input scores to output scores. By embracing such methodologies, which regard tasks as facets of a unified problem, we seek to develop models for symbolic music that not only excel in specific tasks but are also adaptable and proficient across a wide range of tasks.

\begin{figure*}[t]
    \centering
    \includegraphics[width=\textwidth]{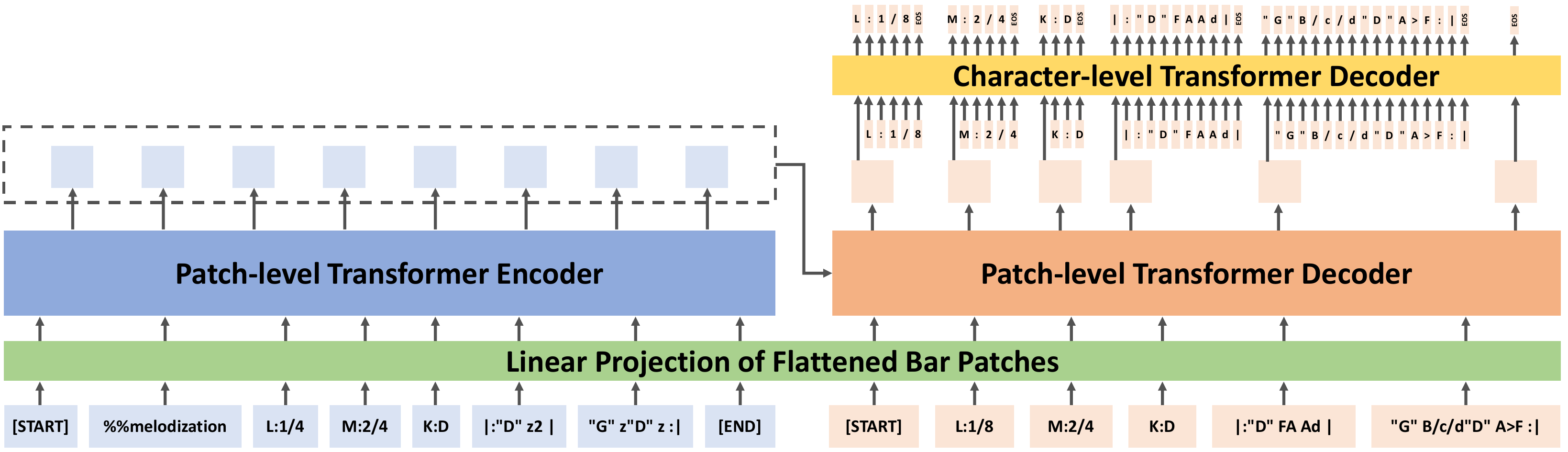}
    \caption{The MelodyT5 framework employs a Transformer encoder-decoder architecture with bar patching for music processing. It uses linear projection of input bar patches, fed into a patch-level Transformer encoder. The encoder output provides context for a patch-level Transformer decoder to autoregressively produce target bar features. A character-level Transformer decoder then uses these features to generate detailed characters for each bar, forming the target musical score.}
    \vspace{-1em}
\end{figure*}

In this paper, we introduce MelodyT5, which leverages an encoder-decoder framework to perform multiple symbolic music tasks as unified score-to-score transformations. Pre-trained on the MelodyHub dataset, which contains over 1 million task instances across seven melody-centric tasks in ABC notation, MelodyT5 overcomes the limitations of task-specific models and sparse data availability in symbolic music processing. By implementing bar patching \cite{DBLP:conf/ismir/WuY0S23, DBLP:conf/hcmir/WuLY023}, MelodyT5 can handle longer sequences effectively, expanding its applicability to a wider range of tasks while maintaining computational efficiency. Our results underscore the promise of employing multi-task learning approaches in symbolic music processing, demonstrating the superior performance of MelodyT5 across a spectrum of tasks and providing a rich dataset for future research. The key contributions of our paper are as follows:

\begin{itemize}
    \item MelodyT5, employing an encoder-decoder approach, redefines symbolic music processing by including multiple tasks as unified score-to-score transformations, demonstrating versatility and breaking traditional task-specific constraints.
    \item MelodyHub, a dataset comprising 261,900 unique melodies in ABC notation across over 1 million task instances for seven different tasks, serves as the cornerstone for effective pre-training of MelodyT5.
    \item Our experimental results demonstrate the efficacy of multi-task transfer learning in symbolic music, with models trained across multiple tasks outperforming those trained in isolation.
\end{itemize}
\section{Methodology}
In this section, we delve into the methodology behind MelodyT5. We first introduce the ABC notation and bar patching for music representation, then present the architectural design of MelodyT5, and finally outline the pre-training objective of our model, which focuses on score-to-score transformations as the basis for multi-task learning.

\subsection{Data Representation}
We utilize ABC notation, a concise symbolic music format, for encoding musical scores with ASCII characters. This text-based format elegantly represents musical elements like notes, rhythms, and articulations in a human-readable manner, thereby facilitating thorough music documentation. Additionally, it promotes the utilization of NLP techniques for both music analysis and generation, as evidenced by recent studies \cite{wu2023exploring, casini2023generating, DBLP:journals/corr/abs-2402-16153}.

To process musical scores encoded in ABC notation more efficiently, we implement the bar patching technique \cite{DBLP:conf/ismir/WuY0S23, DBLP:conf/hcmir/WuLY023}. Bar patching involves breaking down musical sequences into units called bar patches. Each of these units corresponds to either a bar or an information field (such as key and meter), including a sequence of characters that represent musical symbols within that patch. Unlike the conventional character-level or token-level tokenization of ABC notation \cite{DBLP:journals/corr/SturmSBK16, Geerlings2020InteractingWG}, where individual characters or tokens are processed independently, bar patching groups multiple characters into cohesive semantic units. Typically, each patch comprises 10 or more tokens, thus effectively reducing the overall sequence length of musical scores.

\subsection{Model Architecture}
As shown in Fig. 1, the MelodyT5 framework employs an encoder-decoder architecture based on the Transformer network \cite{DBLP:conf/nips/VaswaniSPUJGKP17}, tailored for symbolic music processing. Integrating bar patching into MelodyT5 requires the incorporation of two additional components: a linear projection layer and a character-level decoder. Consequently, the model architecture encompasses the following modules:

\begin{table*}[t!]
  \caption{The MelodyHub collection statistics include the number of instances for each task along with the corresponding data sources. The JSB Chorales dataset is augmented to 15 keys due to its small size and original data in the C key.}
  \vspace{1em}
  \centering
  \resizebox{\linewidth}{!}{%
  \begin{tabular}{lccccccc}
      \toprule
      \textbf{Data Sources} & \textbf{Cataloging} & \textbf{Generation} & \textbf{Harmonization} & \textbf{Melodization} & \textbf{Segmentation} & \textbf{Transcription} & \textbf{Variation} \\
      \midrule
      \textit{ABC Notation} \cite{abcnotation}       & 184,660              & 184,738              & 31,732                  & 31,690                  & ——                     & 174,779                 & ——                 \\
      \textit{FolkWiki} \cite{folkwiki}          & 6,610                & 6,767                & 1,207                   & 1,205                   & ——                     & 6,218                   & ——                 \\
      \textit{JSB Chorales} \cite{deepchoir}      & 4,980                & 4,980                & 4,950                   & 4,950                   & 19,125                  & 4,980                   & ——                 \\
      \textit{KernScores} \cite{kernscores}        & 1,731                & 1,776                & ——                     & ——                     & 1,275                   & 1,754                   & ——                 \\
      \textit{Meertens Tune Collections} \cite{liederenbankmtc} & 16,662         & 16,662               & ——                     & ——                     & 16,660                  & 16,297                  & ——                 \\
      \textit{Nottingham} \cite{nottingham}        & 1,031                & 1,031                & 1,014                   & 1,014                   & ——                     & 1,021                   & ——                 \\
      \textit{OpenScore Lieder} \cite{openscoreliedercorpus} & 1,326                & 1,326                & ——                     & ——                     & ——                     & 1,255                   & ——                 \\
      \textit{The Session} \cite{thesession}      & 44,620               & 44,620               & 3,081                   & 3,078                   & ——                     & 42,838                  & 174,104             \\
      \midrule
      \textit{Total}             & 261,620              & 261,900              & 41,984                  & 41,937                  & 37,060                  & 249,142                 & 174,104              \\
      \bottomrule
  \end{tabular}%
  }
\vspace{-1em}
\end{table*}

\textbf{Linear Projection:} This component converts each bar patch into a dense embedding. It takes a multi-hot vector as input, formed by concatenating one-hot vectors representing characters within the bar patch with the shape \( S \times V \), where \( S \) represents the patch size (i.e., the maximum number of characters in a patch) and \( V \) represents the vocabulary size. If a patch contains fewer than \( S \) characters, it will be padded with a special token to make it a \( S \)-character patch. The vector is then mapped to a dense embedding, serving as input to the patch-level encoder or decoder.

\textbf{Patch-level Encoder:} It is responsible for generating contextualized representations to understand the input musical score by operating on the dense embeddings produced by the linear projection layer. Leveraging mechanisms like self-attention and feed-forward neural networks, it captures global dependencies within the input musical score.

\textbf{Patch-level Decoder:} Tasked with generating the dense representation of the next bar patch, the patch-level decoder utilizes contextualized representations from the encoder and patch embeddings of previously generated content. It employs cross-attention and autoregressive generation mechanisms, ensuring global coherence and continuity in the sequence of generated bar patches.

\textbf{Character-level Decoder:} Operating in a step-by-step manner, the character-level decoder produces characters within the next bar patch based on the dense representation generated by the patch-level decoder. By utilizing the dense representation as a context vector, it decodes each character within the bar, focusing on local information, and sequentially reconstructs every bar patch until it completes the generation of the target musical score.

The encoder-decoder architecture with bar patching in MelodyT5 enables efficient score-to-score transformations by hierarchically modelling music at both patch and character levels, capturing global structure and local details inherent in compositions.

\subsection{Pre-training Objective}
The pre-training objective of MelodyT5 aims to optimize a unified encoder-decoder framework for processing and generating symbolic music across a variety of tasks, utilizing cross-entropy loss for next token prediction.

We consider a dataset \(D\) consisting of pairs \((X, Y)\), where \(X\) is an input musical score and \(Y\) is the target musical score. Each score is represented as a sequence of bar patches \( \{B_1, B_2, \ldots, B_n\} \), with each bar patch \(B_i\) further decomposed into a sequence of characters \( \{c_1, c_2, \ldots, c_m\} \). The model is trained to predict each token (i.e., character) of the target score given the input score and the previously generated tokens in an autoregressive manner.

Formally, the pre-training objective can be represented as minimizing the cross-entropy loss across all tokens in the target sequence:
\begin{equation}
\mathcal{L}(\theta) = -\sum_{(X,Y) \in D} \sum_{i=1}^{n} \sum_{j=1}^{m} \log P_\theta(c_j^i | X, B_{<i}, c_{<j}^i)
\end{equation}
where \(c_j^i\) is the \(j\)-th character in the \(i\)-th bar patch of score \(Y\), \(B_{<i}\) includes all bar patches before the \(i\)-th, \(c_{<j}^i\) are characters before the \(j\)-th in the current patch, and \(P_\theta\) is the probability of the model, parameterized by \(\theta\), of predicting the correct character.

This objective incorporates the fundamental principle that the vast majority of symbolic music tasks can be considered as transformations from score to score, or, in other words, from an input musical score to a target musical score. By pre-training on this objective, MelodyT5 acquires the ability to understand and replicate a wide array of patterns and structures inherent to different music tasks, which is pivotal for its success across various applications within symbolic music processing.

\section{Dataset}
This section outlines the melody curation and task definition of the MelodyHub dataset. MelodyHub, crucial for training MelodyT5, comprises seven melody-centric tasks. This collection, sourced from sheet music datasets, includes folk songs and other non-copyrighted musical scores from various traditions and epochs.

\subsection{Melody Curation}
The MelodyHub dataset was curated using publicly available sheet music datasets and online platforms, with original formats like ABC notation, MusicXML, and Humdrum. The data curation process included several steps:

\begin{enumerate}
    \item Entries featuring explicit copyright indicators such as ``copyright'' or ``©'' symbols were excluded.
    \item All data was converted to MusicXML format for standardization and subsequently transformed into ABC notation to ensure format consistency.
    \item Melodies consisting of fewer than eight bars were omitted from the dataset to maintain adequate complexity and musical richness.
    \item Removal of lyrics and non-musical content (e.g., contact information of transcribers and URL links) aimed to focus solely on musical notation.
    \item Leading and trailing bars of complete rest were removed from each piece.
    \item Each piece underwent verification for the presence of a final barline, with addition if absent.
    \item Entries were deduplicated to prevent redundancy.
\end{enumerate}

By ensuring the quality and consistency of the MelodyHub dataset, these steps led to a substantial collection of 261,900 melodies with uniform formatting, making it suitable for training and evaluating symbolic music models like MelodyT5.

\subsection{Task Definition}
Following the curation of melody data, the MelodyHub dataset was segmented into seven tasks, as summarized in Table 1, presented in a score-to-score format with input-output pairs. In MelodyHub, every input data includes a task identifier (e.g., \texttt{\%\%harmonization}) at the outset to specify the intended task. Below are the definitions of these tasks:

\textbf{Cataloging:} This task selects melodies with music-related metadata like titles, composers, and geographical origins (e.g., \texttt{C:J.S. Bach}, \texttt{O:Germany}). The input data includes information fields with these attributes, while specific information is removed and the order is randomized. The output includes the corresponding metadata without the musical score.

\textbf{Generation:} Here, the input solely consists of a task identifier (i.e., \texttt{\%\%generation}), while the output comprises comprehensive musical scores. Following TunesFormer \cite{DBLP:conf/hcmir/WuLY023}, control codes are affixed to all melodies as information fields to denote musical structure information. These codes, namely \texttt{S:}, \texttt{B:}, and \texttt{E:}, signify the number of sections, bars per section, and edit distance similarity between every pair of sections within the tune.

\textbf{Harmonization:} This task involves melodies containing chord symbols. The chord symbols are removed from the input, while the original data is retained as the output. An additional information field denoting edit distance similarity (\texttt{E:}) is appended to the output, indicating the similarity between the input and output, ranging from 0 to 10 (no match at all to exact match). Lower similarity values suggest the need for more chord symbols.

\textbf{Melodization:} In contrast to harmonization, this task operates inversely and also employs melodies containing chord symbols. The notes in the original score are replaced with rests, and adjacent rest durations are combined. The resultant score, comprising rests and chord symbols, serves as the input. Similar to harmonization, an \texttt{E:} field is added at the outset of the output, with lower values facilitating the generation of more intricate melodies.

\textbf{Segmentation:} Melodies in Humdrum format (i.e., KernScores and Meertens Tune Collections) containing curly braces indicating segmentation or voices from the JSB Chorales dataset (four-part compositions) with fermatas are chosen. These markers are transformed into breath marks. The input data omits all breath marks, while the output introduces an \texttt{E:} field at the beginning to aid the generation of breath marks, with lower values implying the need for more breath marks to be added.

\textbf{Transcription:} ABC notation is initially converted to MIDI, then reconverted back to ABC. The resultant ABC from the MIDI conversion loses substantial score information, such as distinguishing enharmonic equivalents and missing musical ornaments (e.g., trill). The MIDI-converted ABC serves as the input, while the original ABC, appended with an added \texttt{E:} field, constitutes the output. Lower \texttt{E:} values denote greater discrepancies between the transcribed and input scores, particularly due to absent repeat symbols.

\textbf{Variation:} This task centres on data from The Session, wherein each ABC notation file may contain multiple variants of the same tune. Tunes with two or more variations are selected, with every possible pair of variants utilized as both input and output. The output initiates with an \texttt{E:} field signifying the extent of disparities between the input and output scores, with lower values suggesting substantial variations in the musical scores.

Together, resulting in 1,067,747 task instances in total, these tasks include various MIR challenges from analytical to generative, providing a comprehensive resource\footnote{https://huggingface.co/datasets/sander-wood/melodyhub} for developing symbolic music models like MelodyT5.

\section{Experiments}
This section evaluates the effectiveness of MelodyT5 in symbolic music processing through a series of experiments. It outlines experimental settings, conducts ablation studies on multi-task learning impact, and compares MelodyT5 with baseline models in various tasks.

\subsection{Settings}
The experiments are structured to systematically assess the capabilities of MelodyT5 for diverse symbolic music tasks. We utilize the MelodyHub dataset, which is randomly split into 99\% for training and 1\% for validation.

MelodyT5 features a 9-layer patch-level encoder and decoder with shared weights, a 3-layer character-level decoder, and a hidden size of 768, amounting to 113 million parameters. This configuration processes ABC sequences up to 16,384 characters, with a 256 patch length and a 64 patch size. It employs a 128-size ASCII-based vocabulary, using characters 0-2 for special tokens (pad, bos, and eos).

The AdamW optimizer \cite{DBLP:conf/iclr/LoshchilovH19} is used, setting a learning rate of 2e-4. The process includes a 3-epoch warmup, a constant learning rate over 32 epochs, and a batch size of 10 for each GPU, ensuring consistency in hyperparameter settings across all tasks. It took approximately 2 days to complete the pre-training using 6 RTX 3090 GPUs.

In ablation studies, we investigate the effects of multi-task learning on MelodyT5, considering three settings: 1) omitting pre-training, 2) using only the downstream task-specific data from MelodyHub, or 3) utilizing the entire MelodyHub dataset, which includes all tasks.

In terms of comparative evaluations, we select open-source models that excel in their respective domains for benchmarking. MelodyT5 is fine-tuned on identical datasets to these models, ensuring fairness in comparison. For models trained on proprietary datasets, we retrain them using accessible datasets to ensure reproducibility.

Our objective evaluation strategy includes ablation studies focused on bits-per-byte (BPB) for consistent measurement, alongside task-specific metrics for comparative evaluations. Additionally, A/B tests are conducted for the subjective evaluation against baseline models.

\begin{table*}[htbp]
  \centering
  \caption{Experimental results from ablation studies illustrate the impact of multi-task learning on diverse symbolic music tasks, evaluated through BPB (bits-per-byte) to compare performance across various pre-training settings.}
  \vspace{1em}
  \resizebox{\textwidth}{!}{%
  \begin{tabular}{l|ccccccc}
    \toprule
    \multirow{2}{*}{\textbf{Pre-training}} & \textbf{Cataloging} & \textbf{Generation} & \textbf{Harmonization} & \textbf{Melodization} & \textbf{Segmentation} & \textbf{Transcription} & \textbf{Variation} \\
    & \textit{WikiMT \cite{wikimusictext}} & \textit{Wikifonia \cite{wikifonia_archive}} & \textit{CMD \cite{chord_melody_dataset}} & \textit{EWLD \cite{openewld}} & \textit{Essen \cite{kern_essen}} & \textit{Liederschatz \cite{humdrum_erk}} & \textit{The Session \cite{thesession}} \\
    \midrule
    \textit{None} & 0.0376 & 1.2382 & 0.5680 & 0.7949 & 0.0272 & 1.1938 & 0.4932 \\
    \textit{Task-Specific} & 0.0379 & 0.8850 & 0.3393 & 0.6322 & 0.0224 & 0.3432 & -- \\
    \textit{Multi-Task} & \textbf{0.0350} & \textbf{0.8472} & \textbf{0.2925} & \textbf{0.5067} & \textbf{0.0119} & \textbf{0.2969} & \textbf{0.3949} \\
    \bottomrule
  \end{tabular}
  } 
\end{table*}

\subsection{Ablation Studies}
In our ablation studies, MelodyT5 was evaluated on the test sets of various symbolic music benchmarks. Due to the lack of a directly suitable external dataset for the variation task, we chose to evaluate using the validation set of The Session. As a result, there was no task-specific pre-training for the variation task.

The ablation studies aim to explore two aspects: 1) the overall efficacy of pre-training, particularly in the context of multi-task pre-training versus task-specific pre-training, and 2) the extent to which performance gains from multi-task pre-training vary among different tasks, especially considering differences in the available volume of pre-training data across these tasks.

The ablation studies, as depicted in Table 2, show that pre-training is crucial for improving the performance of symbolic music tasks. Models trained with pre-training consistently outperform those without, indicating that pre-training enhances model generalization and performance. Multi-task pre-training is also superior to task-specific pre-training, as models trained with multi-task pre-training show lower BPB scores. This highlights the importance of leveraging multi-task pre-training to effectively capture shared patterns and structures in symbolic music data, enabling MelodyT5 to generalize better to downstream tasks.

Furthermore, it is noteworthy that while multi-task pre-training consistently yields performance gains across most tasks, the extent of improvement varies, which significantly correlates with the volume of task-specific data available for pre-training. Specifically, tasks with less data, such as segmentation and melodization, showcase more substantial performance gains from multi-task learning. On the other hand, tasks with more data, like generation and cataloging, though still benefiting from multi-task pre-training, show relatively smaller improvements. This observation suggests that while multi-task learning enhances model performance across the board, its impact is especially notable in data-constrained scenarios.

In summary, the ablation studies demonstrate the effectiveness of multi-task learning and underscore the impact of data volume on the benefits derived from such an approach. Multi-task learning boosts model performance across symbolic music tasks and provides notable advantages for tasks with limited data by leveraging shared knowledge across tasks.

\subsection{Comparative Evaluations}
For comparative evaluations, we compare MelodyT5, which is multi-task pre-trained on MelodyHub, with several task-specific baseline models, focusing on melody generation, harmonization, melodization, and segmentation. These tasks are well-established and have open-source models as competitive baselines. The following baseline models have been selected for comparison:

\begin{itemize}
\item \textbf{TunesFormer} \cite{DBLP:conf/hcmir/WuLY023} is applied for melody generation, featuring a Transformer-based architecture with bar patching and control codes. This approach aims to refine the efficiency of the generation process and ensure adherence to musical forms.

\item \textbf{STHarm} \cite{DBLP:journals/access/RhyuCKL22} is utilized as the baseline in melody harmonization, employing a Transformer framework to convert melodies into chords. Its primary focus is on creating harmonies that preserve the structural integrity of the original melody.

\item \textbf{CMT} \cite{DBLP:journals/access/ChoiPHJP21} is chosen for melodization, which involves generating melodies based on chord progressions. It employs a phased training approach, conditioning the generation of rhythm and pitch on the chords to produce dynamic and coherent musical outputs.

\item \textbf{Bi-LSTM-CRF} \cite{zhang2021symbolic} is used for melody segmentation, integrating Bi-LSTM and CRF to effectively identify and segment melodic phrases for music structure analysis.
\end{itemize}

\begin{table*}[htbp]
  \centering
  \caption{Comparative objective evaluation of the MelodyT5 model against task-specific baselines across various symbolic music tasks, utilizing task-related metrics previously established. The baselines include TunesFormer \cite{DBLP:conf/hcmir/WuLY023} for generation, STHarm \cite{DBLP:journals/access/RhyuCKL22} for harmonization, CMT \cite{DBLP:journals/access/ChoiPHJP21} for melodization, and Bi-LSTM-CRF \cite{zhang2021symbolic} for segmentation.}
  \vspace{1em}
    \resizebox{0.85\textwidth}{!}{%
    \begin{tabular}{@{\hspace{2mm}}p{1.5cm}|ccccccccc@{}}
    \toprule
    \multirow{2}{*}{\raggedright\textbf{Models}} & \multicolumn{1}{c}{\textbf{Generation}} & \multicolumn{3}{c}{\textbf{Harmonization}} & \multicolumn{3}{c}{\textbf{Melodization}} & \textbf{Segmentation} \\ 
    \cmidrule(r){2-2} \cmidrule(lr){3-5} \cmidrule(lr){6-8} \cmidrule(l){9-9}
     & \textbf{CTRL↑} & \textbf{CTnCTR↑} & \textbf{PCS↑} & \textbf{MCTD↓} & \textbf{CTnCTR↑} & \textbf{PCS↑} & \textbf{MCTD↓} & \textbf{F1 Score↑} \\ 
    \midrule
    \textit{MelodyT5} & \textbf{0.8664} & \textbf{0.7108} & \textbf{0.3274} & \textbf{1.2080} & 0.8438 & \textbf{0.5084} & \textbf{1.0320} & \textbf{0.9055} \\
    \textit{Baselines} & 0.8162 & 0.5963 & 0.2343 & 1.3125 & \textbf{0.8607} & 0.4863 & 1.0610 & 0.8400 \\
    \bottomrule
    \end{tabular}
    }
\end{table*}

For an objective and quantifiable performance assessment that ensures reproducibility, we leverage previously established task-specific metrics. The selected metrics for our assessment include:

\begin{itemize}
\item \textbf{CTRL (Controllability)} \cite{DBLP:conf/hcmir/WuLY023}: Evaluates the precision of generation control through edit distance similarity between intended and actual control codes.

\item \textbf{CTnCTR \& PCS \& MCTD} \cite{DBLP:journals/corr/abs-2001-02360}: These chord/melody harmonicity metrics evaluate harmonization and melodization tasks by assessing harmonic and melodic compatibility between melodies and chords.

\item \textbf{F1 Score}: Measures the balance between precision and recall in identifying correctly segmented melodic phrases.
\end{itemize}

Table 3 shows that MelodyT5 outperforms task-specific baselines in all tasks. It surpasses the specialized baseline TunesFormer in melody generation, demonstrating enhanced control and precision in generating melodies according to specific musical forms. MelodyT5 leads in harmonization, producing chords that are harmonically compatible with the given melodies while maintaining structural coherence. Although slightly trailing CMT in CTnCTR, it still shows robust performance in other metrics for melodization, demonstrating its ability to generate melodies well integrated with chord progressions. Its performance in melody segmentation is significant, indicating its ability to accurately discern and segment melodic phrases. This performance, achieved without task-specific modifications, highlights the effectiveness of multi-task transfer learning combined with unified score-to-score transformations in symbolic music processing.

In addition to the objective metrics presented in Table 3, we recognize the limitations of solely relying on such measures to evaluate the quality of generated music. Thus, we further explored these areas (i.e., generation, harmonization, and melodization) through subjective experiments to capture listener preferences.

For our subjective evaluation, we randomly chose 30 pieces from the test set for each task and conducted blind A/B testing. Participants were presented with one randomly chosen pair from each of these 30 pairs to compare musical scores generated by MelodyT5 and baseline models under identical conditions. They were asked to choose between MelodyT5, the baseline, or no preference. Each comparison was included in two videos, showcasing both the audio and the Sibelius-rendered musical scores.

For generation, we compared the quality of melodies generated by MelodyT5 and TunesFormer, given the same control codes and information fields. In harmonization and melodization, comparisons were made against baselines given identical melodies or chords, respectively. To ensure fairness, especially considering the baseline model for melodization was limited to generating outputs of only 8 bars, we trimmed the MelodyT5-generated scores to match the output length of this baseline model.

The study involved 155 responses from students and educators with music specializations, ensuring deep understanding of melody and harmony. To secure data reliability, submissions were filtered out of those completed in less than half the overall average duration of 4 minutes and 39 seconds, i.e., those under 2 minutes and 20 seconds. This resulted in a final tally of 124 valid questionnaires.

\begin{figure}[t]
    \centering
    \vspace{-1em}
    \includegraphics[width=0.475\textwidth]{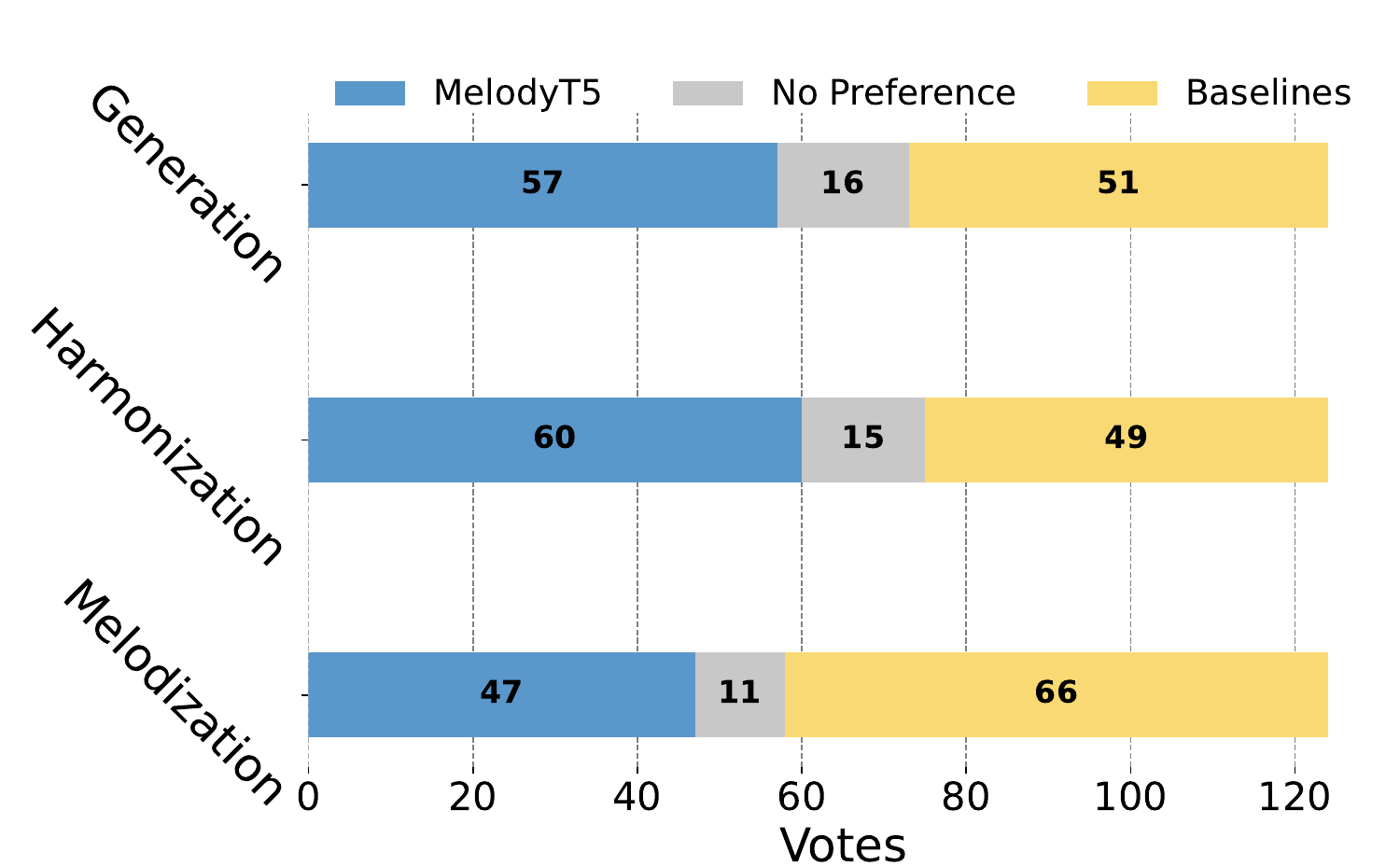}
    \vspace{-2em}
    \caption{Comparative subjective evaluation of MelodyT5 against task-specific baselines in symbolic music tasks, showing vote counts for each model.}
    \vspace{-1em}
\end{figure}

Based on the subjective evaluation in Fig. 2, we observe a notable preference for the MelodyT5 model over the baseline in the tasks of melody generation and harmonization, with MelodyT5 receiving a higher number of votes. However, the preferences reverse in the task of melodization, where the baseline model receives a greater number of votes compared to MelodyT5. This indicates that the baseline model CMT, which employs a two-phase training process focusing separately on rhythm and pitch conditioned on chord progressions, may align more closely with human rhythmic tendencies in melodization, leading to a preference for its outputs in the subjective evaluation. 

Overall, MelodyT5 excels in symbolic music processing, outperforming task-specific models in most tasks and demonstrating the effectiveness of multi-task transfer learning in this domain, despite occasional shortcomings.

\section{Conclusions}

This study presents MelodyT5, a model addressing challenges in symbolic music processing by providing a unified framework for diverse tasks. By treating music tasks as score-to-score transformations, MelodyT5 significantly improves symbolic music processing through multi-task transfer learning. Objective and subjective evaluations demonstrate that MelodyT5 generally outperforms or matches task-specific baseline models without modification. The MelodyHub dataset, with over one million task instances, offers a rich resource for training and evaluating models. While excelling in melody-centric tasks, further optimization is required to tackle more complex musical compositions, such as polyphonic arrangements.



\section{Ethics Statement}

In our research, we are committed to upholding ethical standards regarding data privacy and copyright protection. We diligently adhere to these principles throughout our data collection and processing procedures. All data utilized in our study is sourced from publicly accessible repositories, and we have taken measures to exclude any known copyrighted materials.

\section{Acknowlegdements}
We would like to express our sincere gratitude to Associate Professor Bob L. T. Sturm from KTH Royal Institute of Technology for his valuable discussions and insights into the early stages of the project. We also acknowledge Yuanliang Dong and Jiafeng Liu from the Central Conservatory of Music for their assistance, and Leqi Peng from Fuyin Technology for her support with subjective experiments.

This work was supported by the following funding sources: Special Program of National Natural Science Foundation of China (Grant No. T2341003), Advanced Discipline Construction Project of Beijing Universities, Major Program of National Social Science Fund of China (Grant No. 21ZD19), and the National Culture and Tourism Technological Innovation Engineering Project (Research and Application of 3D Music).

\bibliography{ISMIRtemplate}

%
%
%
%
%

\end{document}